\documentclass[12pt]{iopart}
\usepackage{graphicx}

\begin{document}

\title[Proton-driven plasma wakefield acceleration]
{Proton-driven plasma wakefield acceleration: a path to the future of high-energy particle physics}

\author{AWAKE Collaboration \\
\ \\
R. Assmann$^7$, R. Bingham$^{19,21}$, T. Bohl$^3$, C. Bracco$^3$, \\
B. Buttensch\"on$^{16}$, A. Butterworth$^3$, A. Caldwell$^{16}$,\\ 
S. Chattopadhyay$^{5,11,12,15}$, S. Cipiccia$^{3,21}$, E. Feldbaumer$^3$, \\ 
R.A.  Fonseca$^{6,9}$, B. Goddard$^3$, M. Gross$^{7}$, O. Grulke$^{17}$, \\
E. Gschwendtner$^3$, J. Holloway$^{19,20}$, C. Huang$^{13}$, \\
D. Jaroszynski$^{21}$, S. Jolly$^{20}$, P. Kempkes$^{17}$, N. Lopes$^{9,10}$, \\
K. Lotov$^{2,18}$,  J. Machacek$^{16}$, S.R. Mandry$^{16,20}$, J.W. McKenzie$^1$, \\
M. Meddahi$^3$, B.L. Militsyn$^1$, N. Moschuering$^{14}$, P. Muggli$^{16}$, \\
Z. Najmudin$^{10}$, T.C.Q. Noakes$^1$, P.A. Norreys$^{4,19}$,   E. \"Oz$^{16}$, \\
A. Pardons$^3$, A. Petrenko$^{2,3}$, A. Pukhov$^8$, K. Rieger$^{16}$, \\
O. Reimann$^{16}$, H. Ruhl$^{14}$, E. Shaposhnikova$^3$, L.O. Silva$^9$,  \\
A. Sosedkin$^{2,18}$, R. Tarkeshian$^{16}$, R.M.G.N. Trines$^{19}$, \\
T. T\"uckmantel$^8$, J. Vieira$^{9,16}$, H. Vincke$^3$, M. Wing$^{20}$,  G. Xia$^{5,15}$}
 
\vspace{1cm}
\address{$^1$ Accelerator Science and Technology Centre, ASTeC, STFC Daresbury Laboratory, Warrington, UK}
\address{$^2$ Budker Institute of Nuclear Physics SB RAS, Novosibirsk, Russia}
\address{$^3$ CERN, Geneva, Switzerland}
\address{$^4$ Clarendon Laboratory, University of Oxford, UK}
\address{$^5$ Cockroft Institute, Daresbury, UK}
\address{$^6$ DCTI / ISCTE - Instituto Universit\'{a}rio de Lisboa, Lisboa, Portugal}
\address{$^7$ DESY, Germany}
\address{$^8$ Heinrich Heine University, D\"usseldorf, Germany}
\address{$^9$ GoLP / Instituto de Plasmas e Fusao Nuclear, Instituto Superior Tecnico, Universidade de Lisboa, Lisboa, Portugal}
\address{$^{10}$ John Adams Institute for Accelerator Science, Imperial College, London, UK}
\address{$^{11}$ Department of Physics, University of Lancaster, UK} 
\address{$^{12}$ Department of Physics, University of Liverpool, UK}
\address{$^{13}$ Los Alamos National Laboratory, New Mexico, USA}
\address{$^{14}$ Ludwig Maximilian University, Munich, Germany}
\address{$^{15}$ School of Physics and Astronomy, University of Manchester, UK}
\address{$^{16}$ Max Planck Institute for Physics, Munich, Germany}
\address{$^{17}$ Max Planck Institute for Plasma Physics, EURATOM Association, Greifswald, Germany}
\address{$^{18}$ Novosibirsk State University, Novosibirsk, Russia}
\address{$^{19}$ Rutherford Appleton Laboratory, Chilton, UK}
\address{$^{20}$ University College London, London, UK}
\address{$^{21}$ University of Strathclyde, Glasgow, Scotland, UK}

\ead{m.wing@ucl.ac.uk}
\begin{abstract}
New acceleration technology is mandatory for the future elucidation of fundamental particles and their interactions.  A 
promising approach is to exploit the properties of plasmas. Past research has focused on creating large-amplitude 
plasma waves by injecting an intense laser pulse or an electron bunch into the plasma. However, the maximum 
energy gain of electrons accelerated in a single plasma stage is limited by the energy of the driver. Proton bunches 
are the most promising drivers of wakefields to accelerate electrons to the TeV energy scale in a single stage. An 
experimental program at CERN -- the AWAKE experiment -- has been launched to study in detail the important 
physical processes and to demonstrate the power of proton-driven plasma wakefield acceleration. Here we review 
the physical principles and some experimental considerations for a future proton-driven plasma wakefield accelerator.
\end{abstract}

\maketitle

\section{Introduction}

Over the last fifty years, accelerators of ever increasing energy have been used to probe the fundamental structure 
of the physical world. This has culminated so far in the Large Hadron Collider (LHC) at CERN, Geneva, an accelerator of 
27\,km in circumference. With this accelerator, the Higgs Boson, the particle of the Standard Model that attributes to 
particles their mass, was recently discovered~\cite{ATLAS,CMS} in proton--proton collisions. However, although the 
Standard Model has been incredibly successful at describing fundamental particles and the forces that act between 
them, there are still several unexplained phenomena that pose some of the big questions in science: 

\begin{itemize}

\item Why are the masses of the fundamental particles so different, e.g.\ the top quark and neutrinos? 

\item Why are there three families of quarks and leptons? 

\item Where is the anti-matter in the Universe? 

\item Why does the visible matter constitute only 5\% of the Universe and what are dark matter and dark energy that 
constitute the rest? 

\item Is there a Grand Unification Theory that merges the fundamental forces into one? 

\end{itemize}

That the Standard Model can not answer all these questions, points towards the need for new theories or phenomena 
such as Supersymmetry, which unifies the forces at high energies and provides a candidate for dark matter, or extra 
spatial dimensions, such as required by string theory. Such phenomena are being searched for at high energy using 
the LHC and any successor. It is widely held that a next energy frontier accelerator should collide electrons and 
positrons at around the Tera-electron-Volts (TeV) energy scale. As electrons and positrons are point-like, fundamental 
objects and the centre-of-mass energy is controlled, a significantly cleaner environment can be achieved than at the 
LHC that collides protons. Such a future electron--positron collider would therefore have the potential to search for new 
physics as well as being able to measure to high precision new phenomena discovered already at the LHC. 

The gradient at which charged particles can be accelerated using today's radio-frequency (RF) or microwave 
technology is limited to about 100\,MeV/m by RF breakdown on and fatigue of the cavity walls.  To reach the TeV scale 
in a linear accelerator, the length of the machine is therefore tens of kilometres.  Circular electron colliders are feasible 
at these energies only at the 100\,km scale due to limitations imposed by synchrotron radiation~\cite{tlep}. At these 
scales it becomes difficult to find a suitable stable geological site and the construction cost of such a machine is 
estimated to be in the range of ten(s) of billions of Euros. Therefore, a new high-gradient accelerator technology must 
be developed to ensure that the energy frontier in particle physics can be investigated experimentally within affordable 
cost, time-scale and space constraints. 

Ionized gases, or plasmas, with densities over a thousand times lower than that of the atmosphere, can sustain 
accelerating gradients (several 10\,GeV/m) orders of magnitude larger than RF structures~\cite{dawson}. These large 
fields are due to the collective response of the plasma electrons to the electric field of a laser pulse or charged particle 
bunch driver.  The plasma, without initial structure, supports waves or wakes travelling at velocities near the speed of 
light, ideal to accelerate particles to relativistic energies. These wakefields have a longitudinal component able to 
extract energy 
from the driver and transfer it to a trailing witness bunch (of e.g.\ electrons). The wakes also have transverse wakefield 
components with focusing strength orders of magnitude larger than that of conventional magnets, allowing for the two 
beams to remain transversely small over long distances. This combination of large plasma fields and long confined 
propagation distances can lead to the large energy gain necessary for high-energy physics applications, but over 
much shorter distances than with today's RF and magnet technology. 

The potential of plasma as the medium for high gradient acceleration has been demonstrated with short and intense 
laser pulse drivers yielding electron bunches of up to 2\,GeV energy gain in cm-long channels~\cite{leemans,wang} 
that corresponds to about 100\,GV/m average accelerating fields.  High gradients have also been demonstrated with 
a short, high charge electron bunch driver with an energy gain of 42\,GeV in 85\,cm, corresponding to 
52\,GV/m~\cite{slac}.

However, in both of these pioneering experiments the energy gain was limited by the energy carried by the driver 
($\sim 100$\,J) and the propagation length of the driver in the plasma ($<1$\,m). The laser pulse and electron bunch driver 
schemes therefore require staging~\cite{staging,leemans-stage}, i.e.\ the stacking of many $10-25$\,GeV acceleration,  
or plasma, stages to reach the $\sim 1$\,TeV energy per particle or equivalently $\sim 2$\,kJ of energy in 
$\sim 2 \times 10^{10}$ electrons and positrons. The scheme proposed in this paper solves these propagation-length and 
energy limitations by using a proton bunch to drive the wakefields. 

Bunches with $3 \times 10^{11}$\,protons and 19\,kJ of energy (the CERN SPS 400\,GeV beam), and with 
$1.7 \times 10^{11}$\,protons and 110\,kJ of 
energy (the CERN LHC 4\,TeV beam) are produced routinely today. Because of their high energy and mass, proton 
bunches can drive wakefields over much longer plasma lengths than other drivers. They can take a witness bunch to 
the energy frontier in a single plasma stage, as was demonstrated in simulations~\cite{caldwell-nphys}. This 
proton-driven scheme therefore greatly simplifies and shortens the accelerator. In addition, because there is no gap 
between the accelerator stages, this scheme avoids gradient dilution.

\section{Self-modulation instability of particle beams in plasmas}

Despite the great potential of proton-driven plasma wakefield acceleration, there are a number of challenges to 
overcome. The major challenge is the length of the existing proton bunches.  A plasma can be understood as an 
ensemble of oscillators swinging at the plasma frequency. To enforce the resonant swinging of these oscillators, 
the driver must contain a Fourier component close to the plasma frequency. The maximum field of a plasma wake 
scales as

\begin{equation}
E_{\rm max} \approx \sqrt{\frac{n_e [{\rm cm}^{-3}]}{10^{14}}} \, {\rm GV/m},
\end{equation}
where $n_e$ is the plasma electron density. Consequently, plasma densities of at least $n_e \approx 10^{14}\,$cm$^{-3}$ 
are required to reach accelerating gradients of GeV/m and above. The corresponding plasma wavelength is

\begin{equation}
\lambda_p \approx \sqrt{\frac{10^{15}}{n_e [{\rm cm}^{-3}]}} \, {\rm mm}.
\end{equation}
At these densities, the plasma wavelength is of the order of a millimetre. On the other hand, proton bunches available 
today are much longer, $\sigma_z = 3-12$\,cm. Having a nearly Gaussian shape, they are not resonant and thus no 
strong plasma wakefields can be excited directly. 

\begin{figure}
\includegraphics[width=\textwidth]{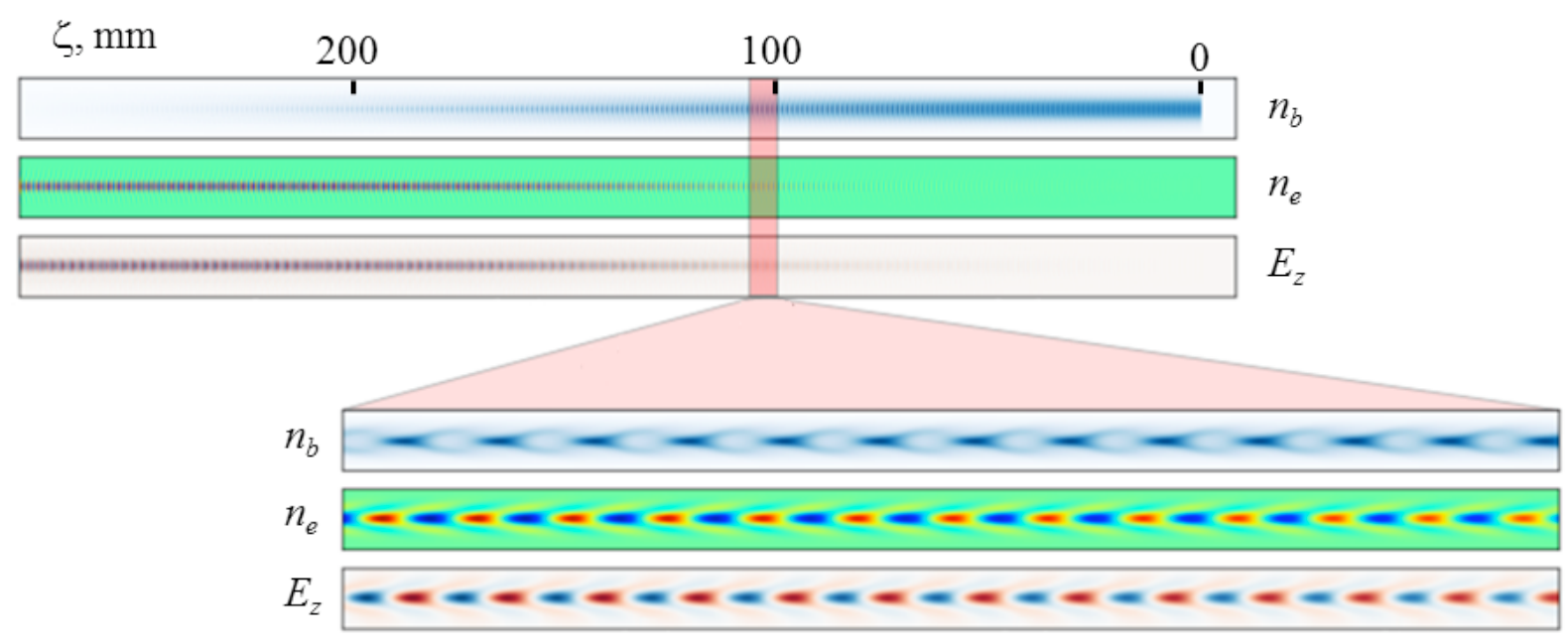}
\caption{Proton beam density $n_b$, plasma electron density $n_e$, and the longitudinal component of the wakefield 
$E_z$ after 4\,m of propagation in plasma. The coordinate $\zeta = ct - z$ is counted from the bunch head. The region 
10\,cm behind the bunch head is zoomed in.}
\label{fig:smi}
\end{figure}

Fortunately, a mechanism has been discovered that automatically splits the proton bunch propagating in plasma 
into a number of micro-bunches: the self-modulation instability (SMI)~\cite{smi-lotov,smi-pukhov}. The instability starts 
from a seeding wave whose transverse field acts on the beam and modulates its radius. The modulation has a period 
very close to the plasma wavelength. Its amplitude grows exponentially from head to tail of the bunch and along the 
propagation distance.  The physics of the instability is now well understood, and theoretical predictions agree well 
with results of simulations~\cite{smi-sat1,smi-sat2,PoP18-024501,lhc,PoP19-063105}.  At saturation, the initially long 
and smooth beam is split into a train of micro-bunches that resonantly excite a strong plasma wave.  This plasma wave 
is inherently weakly nonlinear~\cite{PoP20-083119}, so the way of its excitation is almost independent of the charge 
sign of the drive beam.  An example of a self-modulated bunch as observed in 3D simulations using 
the particle-in-cell code VLPL~\cite{vlpl1,vlpl2} is shown in Figure~\ref{fig:smi}.  Numerical simulation of a 
self-modulating proton beam in the real geometry is a challenging problem, e.g.\ the resolution must be carefully 
chosen~\cite{IPAC13-1238}.

\section{Uniform-density plasma cell}

The second challenge also arises because of a large disparity between the bunch length and the plasma 
wavelength. The proton bunch must be split into some 100\,micro-bunches to drive the high amplitude wake. All 
these micro-bunches must work constructively. This translates into a very strict requirement on the plasma density 
uniformity~\cite{lotov-pukhov-caldwell}. Simulations show that electron trapping and subsequent stable acceleration 
are mainly affected by density gradients. The relative plasma density variations must be controlled down to 

\begin{equation}
\frac{\delta n_e}{n_e} \approx \frac{\lambda_p}{2 \pi \sigma_z}.
\end{equation}
This means that for realistic proton bunches we have to control the plasma density below 0.5\% over distances of 
many metres~\cite{oz-muggli}. 

The best option to achieve the high density uniformity is to fill an evacuated vessel with a neutral gas and ionize it 
instantaneously with a laser. The laser pulse must be short, shorter than $\lambda_p$. If the laser co-propagates 
together with the proton bunch, the fast creation of plasma inside the bunch has the same effect as a sharp 
leading edge of the bunch would have; it reliably seeds the self-modulation.

\section{Injection and acceleration of the witness beam}

The third challenge is a detailed understanding of the interactions of the proton bunch with the plasma to optimize 
the injection capture efficiency (maximize the size of the stable longitudinally focusing and accelerating 
phase-space ``bucket") and properties (minimize the phase space volume) of the accelerated electrons. Theory 
and simulations show that the witness beam should not be injected before the self-modulation instability reaches 
saturation~\cite{smi-sat1,smi-sat2}. The reason is the low phase velocity of the wake during the self-modulation 
linear growth stage. One might split the plasma cell in two parts: the proton bunch would self-modulate in the 
first part and witness particles could be injected into the second, accelerating plasma cell. Yet, injecting the 
witness bunch axially into the second plasma cell is technologically difficult. It is expected that the plasma density 
will not be uniform in the first centimetres behind the cell entry. The irregular wake in the plasma density gradient 
can easily scatter low energy ($10-20$\,MeV) electrons foreseen from an injector. 

\begin{figure}
\includegraphics[width=\textwidth]{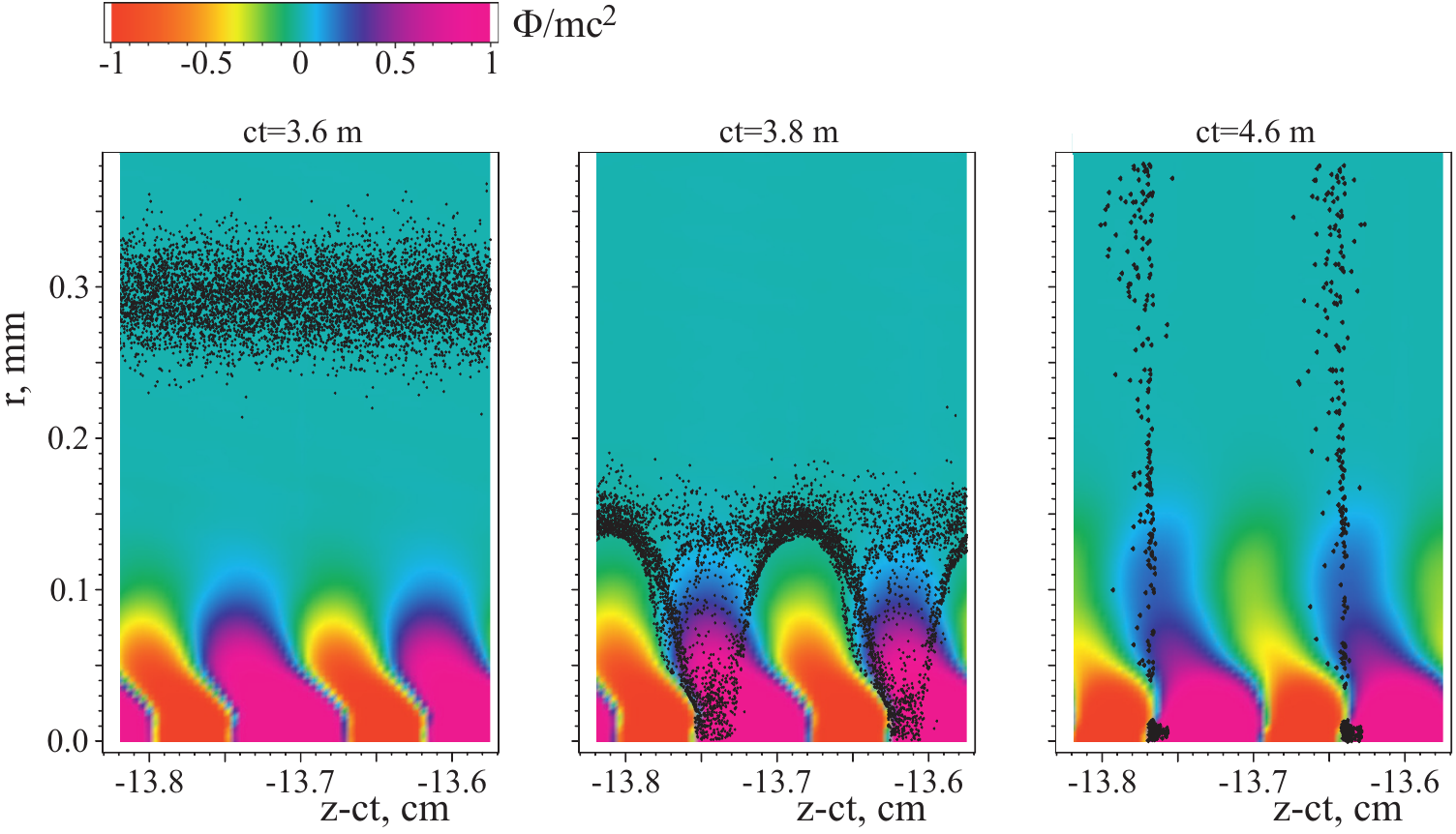}
\caption{Maps of the wakefield potential at three positions along the plasma.  The relative size of the wakefield 
potential is indicated by the colour map.  At $z = 3.6$\,m (left image), the low 
energy electrons (shown as black dots) are injected from the side, at an angle of 9\,mrad, (top on the figure) towards the wakefield.  At 
$z = 3.8$\,m (middle image) the electrons have reached the wakefield potential wells, some are reflected while 
some reach the axis and can be trapped.  At $z = 4.6$\,m (right image), two trapped electron micro-bunches are 
visible near the axis ($r = 0$) and a few electrons are still drifting out radially.}
\label{fig:side}
\end{figure}

The solution is the side injection of particles into the wake of an already modulated bunch~\cite{smi-sat1,side}. In 
this case, the witness electrons propagate at a small angle with respect to the driver and are gradually ``sucked-in" 
at the right phase by the wake's transverse fields (see Figure~\ref{fig:side}). Simulations show that this leads to high quality 
quasi-monoenergetic acceleration of electrons. 

To maximize the capture efficiency of externally injected electrons into the stable longitudinally focusing 
accelerating bucket, it will be necessary to have an electron injector with the flexibility of tuning the injection 
energy from 5\,MeV to 20\,MeV, as well as the capability to change the phase-space volume of the ``cold" electron 
source (charge and emittance) in order to understand the dynamics of the plasma wakefield channel, electron 
capture and acceleration. Such flexibility is offered by a radio-frequency electron gun (RF gun) fitted with a 
laser-driven photocathode as the emitter. 

\section{AWAKE experiment at CERN}

\begin{figure}
\includegraphics[width=\textwidth]{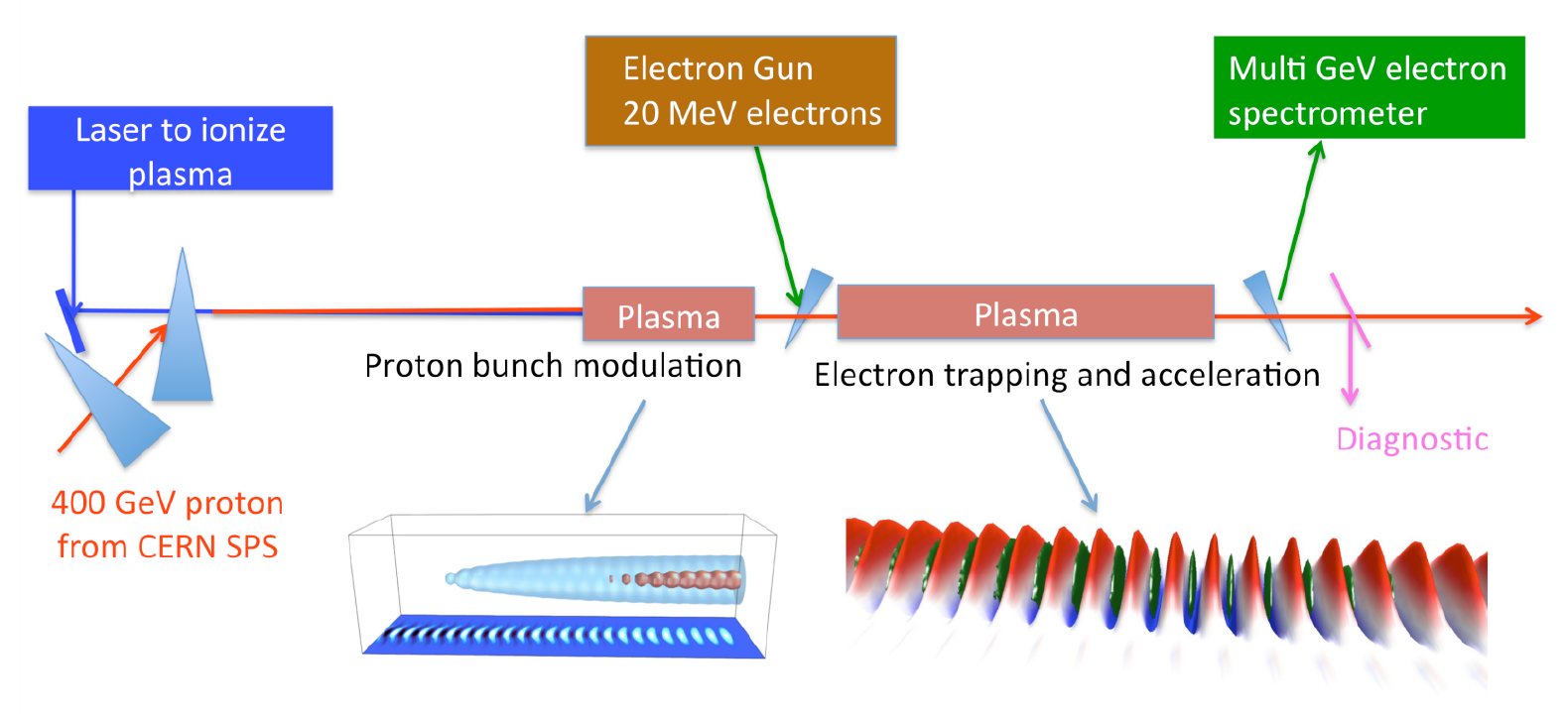}
\caption{Conceptual design of the AWAKE experiment, showing the major sections and description of expected 
effects.}
\label{fig:schematic}
\end{figure}

To address these challenges, the AWAKE experiment at CERN~\cite{tdr} will be the first proton-driven plasma 
wakefield experiment world-wide. The conceptual design of the proposed AWAKE experiment is shown in 
Figure~\ref{fig:schematic}:  The laser and proton bunches are made co-linear. The laser ionizes the metal vapor 
in the first plasma section and seeds the self-modulation instability. The self-modulated proton bunch (shown 
in the left hand side inset) enters a second plasma section where it drives the plasma wakefield structure (shown 
in the right side inset). The electrons are injected in the wakefields and their multi-GeV energy is measured with 
an electron spectrometer. 

\begin{figure}
\includegraphics[width=\textwidth]{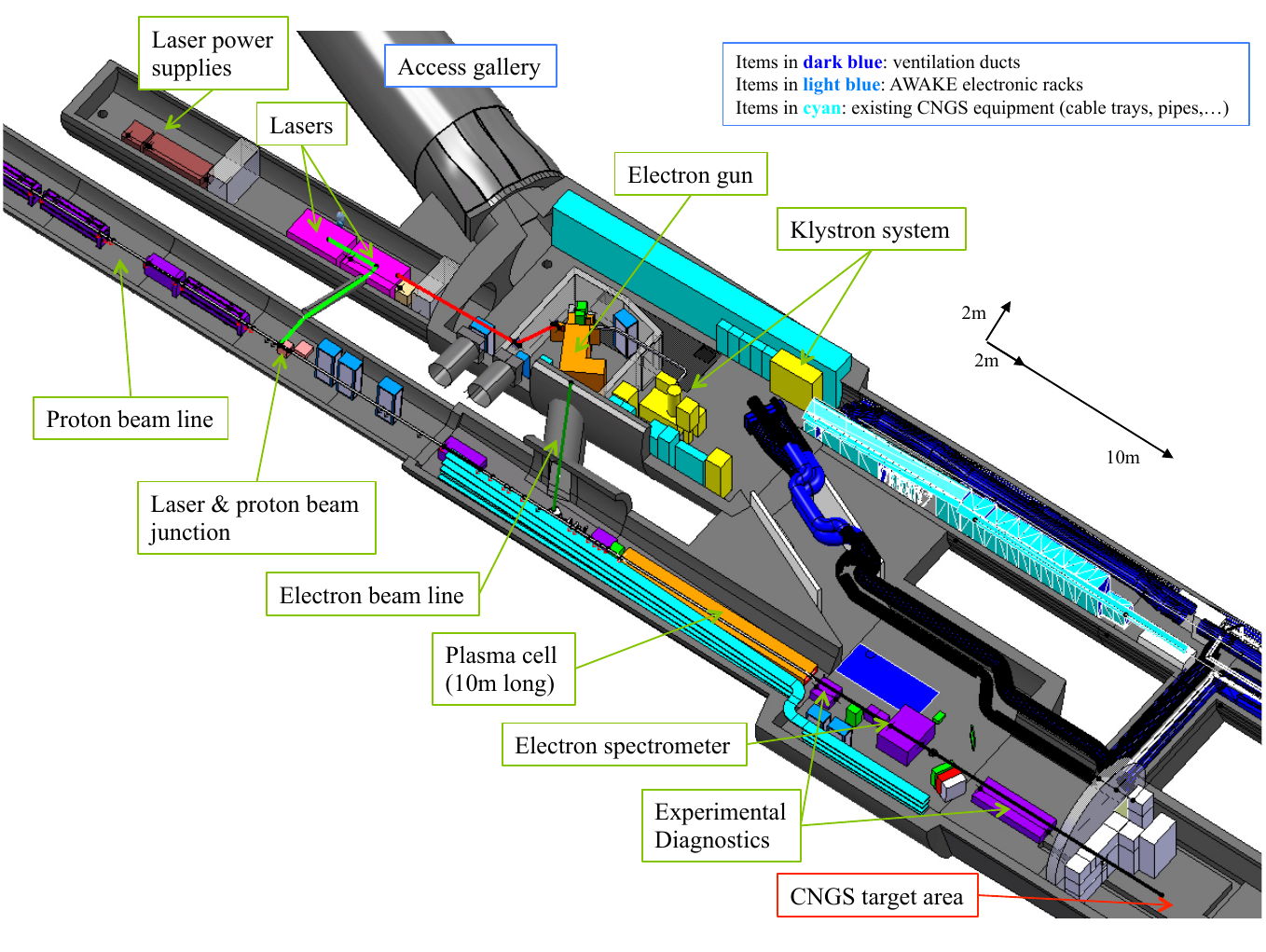}
\caption{Design of the layout of the AWAKE experiment.}
\label{fig:design}
\end{figure}

The AWAKE experiment will be installed in the CERN Neutrinos to Gran Sasso (CNGS) facility~\cite{cngs}. A 
design of the experimental setup is shown in Figure~\ref{fig:design}.  Nominally a proton bunch with intensity 
of the order of $3 \times10^{11}$ protons, a length of 12\,cm and energy of 400\,GeV is extracted every 30\,s 
from the SPS and transported along more than 800\,m of beam line towards the AWAKE experimental area 
that will be installed in the upstream part of the (previous) CNGS facility. The focused transverse beam-size 
at the plasma cell is $\sigma_r = 0.2$\,mm and the transverse normalized emittance is 
$\epsilon_N=3.5$\,mm\,mrad. The laser beam is merged with the proton beam $\sim 20$\,m upstream of the 
entrance of the plasma cell in a junction system. The area downstream of the plasma cell houses the beam 
diagnostics systems and the electron spectrometer. The proton beam will be dumped in the existing CNGS 
hadron stop, $\sim 1,000$\,m downstream of the experimental area, thus avoiding any backscattering of particles 
and radiation into the AWAKE experimental area.

The plasma cell technology which best fulfills the requirements of the experiment is an alkali metal vapor 
source, as alkali metals have been used in previous experiments~\cite{alkali-vapor} and have low ionization 
potentials (e.g.\ 4.2\,eV for the first electron of rubidium). The vapor is therefore relatively easy to ionize with 
a laser pulse, with the threshold intensity being as low as $1.7 \times 10^{12}$\,W/cm$^2$. Rubidium has a large 
ion mass, which makes the plasma less sensitive to ion motion~\cite{ion-motion}.  

\begin{figure}
\begin{center}
\includegraphics[width=0.5\textwidth]{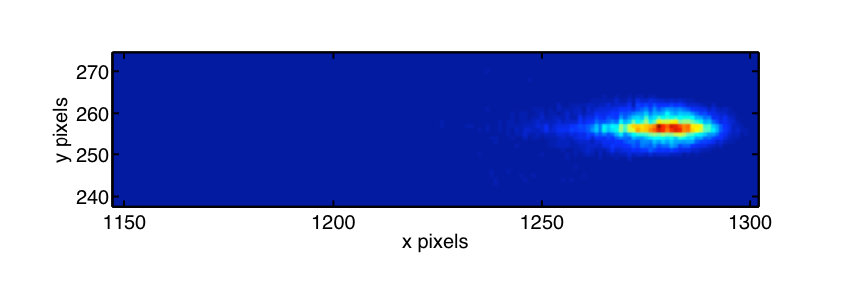}
\put(-222,65){{\bf (a)}}
\end{center}
\includegraphics[width=0.5\textwidth]{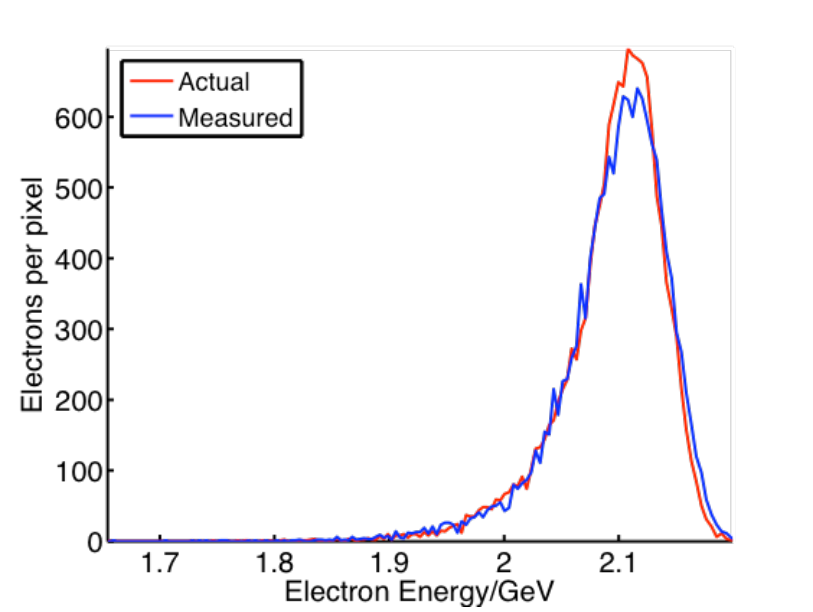}
\includegraphics[width=0.5\textwidth]{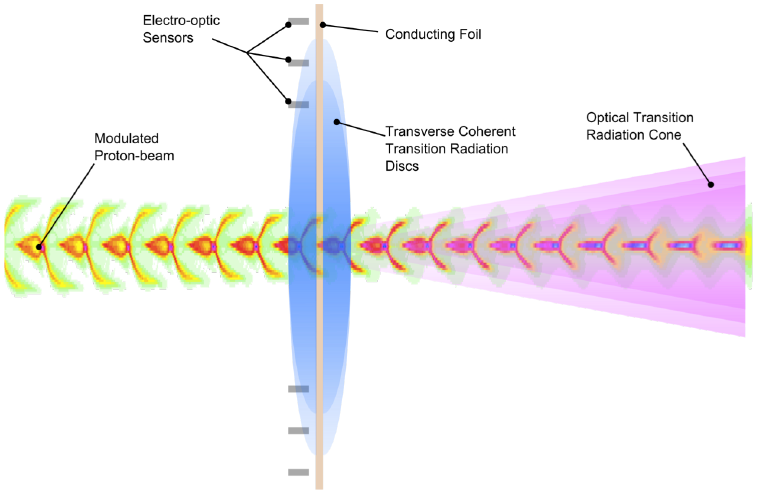}
\put(-448,153){{\bf (b)}}
\put(-228,153){{\bf (c)}}
\caption{a) Simulated impact position of accelerated electrons on a scintillator screen having passed through a 
magnetic spectrometer.  b) The electron energy spectrum reconstructed using the energy spectrometer (blue 
line) compared with the energy distribution exiting the plasma cell (red line).  c) A modulated proton beam passing 
through a conducting foil, leading to transition radiation, detected using a streak camera and electro-optic 
sensors.}
\label{fig:diagnostics}
\end{figure}

Figure~\ref{fig:diagnostics} shows an example of an electron bunch energy spectrum as obtained from 
numerical simulations using the simulation tool LCODE~\cite{lcode} for the side injection case and the 
nominal proton bunch parameters. A simulation of what would be seen on a scintillator screen in the electron 
spectrometer downstream of the plasma cell is also shown as well as the energy reconstructed from this 
spatial spread.  Electron bunches of energy 16\,MeV, charge 0.2\,nC, and length $\sigma_{ze}=2.5$\,mm are 
side injected after 3.9\,m of plasma.  Approximately 5\% of electrons are trapped and accelerated to the end 
of the 10\,m plasma. This amount of charge is not high enough to observe beam loading, but sufficient to 
characterize accelerating gradients.
The final energy spread is $\sim2\%$ r.m.s., which can be accurately reconstructed 
by the electron spectrometer, indicating that percent level energy spread can be reached. The peak gradient 
seen in this simulation is above 1\,GeV/m, and the average gradient witnessed by the electrons is 
350\,MeV/m. 

In addition to the electron spectrometer, several other diagnostic systems will be used to characterize the beams 
and plasma so as to better understand the physics of self-modulation and acceleration of electrons in the wake 
of the proton beam~\cite{reimann-tarkeshian}.  Examples are shown in Figure~\ref{fig:diagnostics}c) in which a proton beam passes through 
a metal foil, thereby producing a cone of optical transition radiation that will be measured using a streak camera.  
Additionally, transverse coherent transition radiation will be produced and detected using electro-optical sensors; 
this will be the first experimental use of this recent concept~\cite{tctr}.

First protons to the experiment are expected at the end of 2016 and this will be followed by an initial $3-4$\,year 
experimental program of four periods of two weeks of data taking.

\section{Towards the TeV Frontier}

The AWAKE experiment at CERN will test the principle of proton beam self-modulation in plasma and electron 
acceleration in the excited wake. However, the future use of proton-driven plasma wakefield acceleration as the 
energy-frontier technique requires additional research effort. The keys here are a scalable plasma source and 
shorter proton bunches. 

Metal vapor plasma sources, ionized with lasers, routinely reach plasma densities of the order of 
$10^{17}$\,cm$^{-3}$ but suffer from limitations of available laser power and are difficult to scale. A solution could 
be a helicon-wave plasma cell or discharge plasma cell that are potentially scalable in length over very long distances. 
These types of plasma cell follow a strictly modular concept although have a number of challenges such as density 
uniformity which are being addressed by an R\&D program.

Shorter proton bunches will allow plasma wakes to be driven with either far fewer micro-bunches (via the SMI) 
or directly with an un-modulated bunch. This would dramatically reduce the uniformity requirements on the 
plasma, simplify the plasma cell technology, and improve the overall energy efficiency of the 
scheme~\cite{caldwell-nphys}.  Various schemes within the current set-up of the CERN accelerators are being 
considered~\cite{compression} as well as compression via magnetic chicanes~\cite{chicane} and the production 
of short proton bunches at source~\cite{short-proton}.

The results of the AWAKE experiment will inform future larger-scale R\&D projects on proton-driven plasma 
wakefield acceleration and could lead to future high energy colliders for particle physics.  Simulations have 
already shown~\cite{lhc} that electrons can be accelerated to the multi-TeV scale (e.g.\ 3\,TeV after 4\,km) 
using the 7\,TeV LHC proton beam which is modulated prior to electron acceleration, following the scheme 
to be investigated by the AWAKE experiment.  An alternative initial application is to provide the electron beam 
for the LHeC project~\cite{LHeC} in which it is planned to collide a 50\,GeV electron beam with the LHC proton 
beam.  The SPS or LHC beams could be used to accelerate electrons using proton-driven plasma wakefield 
acceleration which are then used as the electron beam or as the injector to it. Some of the key issues in 
designing a compact electron--positron linear collider and an electron--proton collider based on existing CERN 
accelerator infrastructure have been identified~\cite{xia-future}.

\section{Outlook}

The AWAKE experiment at CERN will be the first proton-beam driven plasma wake field acceleration experiment 
worldwide. Its success will open a pathway towards a revolutionary plasma-based TeV lepton collider. This 
revolution will then enable groundbreaking particle physics discoveries.

\section*{Acknowledgments}
This work was supported in parts by: BMBF, Germany, project 05H12PF5; EU FP7 EuCARD-2, Grant Agreement 
312453 (WP13, ANAC2); EPSRC and STFC, United Kingdom; and the Ministry of Education and Science of the Russian Federation.
M.~Wing acknowledges the support of DESY, Hamburg.


\section*{References}
\begin{thebibliography}{10}

\bibitem{ATLAS}
G. Aad et al., ATLAS Coll., Phys. Lett. {\bf B~716}, 1 (2012).

\bibitem{CMS}
S. Chatrchyan et al., CMS Coll., Phys. Lett. {\bf B~716}, 30 (2012).

\bibitem{tlep}
M. Koratzinos et al., {\tt http://arxiv.org/pdf/1306.5981.pdf}

\bibitem{dawson}
T. Tajima and J. Dawson, Phys. Rev. Lett. {\bf 43} , 267 (1979).

\bibitem{leemans}
W. Leemans et al., Nature Phys. {\bf 2}, 696 (2006).

\bibitem{wang}
X. Wang et al., Nature Commun. {\bf 4}, 1988 (2013).

\bibitem{slac}
I. Blumenfeld et al., Nature {\bf 445}, 741 (2007)

\bibitem{staging}
S. Chattopadhyay and R.M. Jones, Nucl. Instrum. Meth.  {\bf A~657}, 168 (2011).

\bibitem{leemans-stage}
W. Leemans and E. Esarey, Physics Today, {\bf 62}(3), 44 (2009). 

\bibitem{caldwell-nphys}
A. Caldwell, K. Lotov, A. Pukhov and F. Simon, Nature Phys. {\bf  5}, 363 (2009).

\bibitem{smi-lotov}
K.V. Lotov, Proc. 6th European Particle Accelerator Conference, Stockholm, p.806 (1998).

\bibitem{smi-pukhov}
N. Kumar, A. Pukhov and K. Lotov, Phys. Rev. Lett. {\bf 104}, 255003 (2010). 

\bibitem{smi-sat1}
A. Pukhov et al., Phys. Rev. Lett. {\bf 107}, 145003 (2011).

\bibitem{smi-sat2}
C. Schroeder et al., Phys. Rev. Lett. {\bf 107}, 145002 (2011).

\bibitem{PoP18-024501}
K.V. Lotov, Phys. Plasmas {\bf 18}, 024501 (2011).

\bibitem{lhc}
A. Caldwell and K.V. Lotov, Phys. Plasmas {\bf 18}, 103101 (2011).

\bibitem{PoP19-063105} 
J. Vieira et al., Phys. Plasmas {\bf 19}, 063105 (2012).

\bibitem{PoP20-083119} 
K.V. Lotov, Phys. Plasmas {\bf 20}, 083119 (2013).

\bibitem{vlpl1}
A. Pukhov, J. Plasma Phys. {\bf 61}, 425 (1999).

\bibitem{vlpl2}
T. Tuckmantel et al., Plasma Science, IEEE Transactions on {\bf 38}, 2383 (2010).

\bibitem{IPAC13-1238} 
K.V. Lotov, A. Sosedkin and E.Mesyats, Proc. IPAC2013 (Shanghai, China), p.1238 (2013).

\bibitem{lotov-pukhov-caldwell}
K.V. Lotov, A. Pukhov and A. Caldwell, Phys. Plasmas {\bf 20}, 013102 (2013).

\bibitem{oz-muggli}
E. \"{O}z and P. Muggli, Nucl. Instrum. Meth. {\bf A~740}, 197 (2014).

\bibitem{side}
K.V. Lotov, J. Plasma Phys. {\bf 78}, 455 (2012).

\bibitem{tdr}
A. Caldwell et al., CERN-SPSC-2013-013 (2013).

\bibitem{cngs}
E. Gschwendtner et al., Proc. IPAC2010, Kyoto, p.4164 (2010).

\bibitem{alkali-vapor}
P. Muggli et al., Plasma Science, IEEE Transactions on {\bf 27}, 791 (1999).

\bibitem{ion-motion}
J. Vieira et al., Phys. Rev. Lett. {\bf 109}, 145005 (2012).

\bibitem{lcode}
K.V. Lotov, Phys. Rev. ST Accel. Beams {\bf 6}, 061301 (2003).

\bibitem{reimann-tarkeshian}
O. Reimann and R. Tarkeshian, Proceedings of IBIC2013, p.467 (2013).

\bibitem{tctr}
A. Pukhov and T. Tueckmantel, Phys. Rev. ST Accel. Beams {\bf 15}, 111301 (2012).

\bibitem{compression}
H. Timko et al., Proc. IPAC2013, Shanghai, p.1820 (2013).

\bibitem{chicane}
G. Xia and A. Caldwell, Proc. IPAC2010, Kyoto, p.4395 (2010).

\bibitem{short-proton}
F.L. Zheng et al., Phys. Plasmas {\bf 19}, 023111 (2012).

\bibitem{LHeC}
J.L. Abelleira Fernandez et al., J. Phys. {\bf G~39}, 075001 (2012).

\bibitem{xia-future}
G. Xia et al., Nucl. Instrum. Meth. {\bf A~740}, 26 (2014). 


\end{thebibliography}
\end{document}